\documentclass[a4paper,11pt]{article}
\usepackage{jinstpub} 
\usepackage{lineno}

\usepackage[utf8]{inputenc} 
\usepackage[T1]{fontenc}    
\usepackage{hyperref}       
\usepackage{url}            
\usepackage{booktabs}       
\usepackage{amsfonts}       
\usepackage{nicefrac}       
\usepackage{microtype}      
\usepackage{lipsum}
\usepackage{amsmath}
\usepackage{graphicx}
\usepackage{float}
\usepackage{slashed}
\graphicspath{{Images/}}

\makeatletter
\gdef\@fpheader{}
\makeatother
\begin{document}
\title{\textbf{Probing Dark Matter via Effective Field Theory Approach}}

\author[a]{Ayşe Elçiboğa Kuday}
\emailAdd{ayse.kuday@msgsu.edu.tr}

\author[a]{Ferhat Özok} 
\emailAdd{ferhat.ozok@ccern.ch}
\affiliation[a]{Department of Physics, Mimar Sinan Fine Arts University,  34380, Istanbul, TURKEY}  

\author[b]{Erdinç Ulaş Saka}
\affiliation[b]{İstanbul University Faculty of Science Department of Physics, 34134 , Istanbul, TURKEY}
\emailAdd{ulassaka@istanbul.edu.tr}

\abstract{
We analyse dark matter in most general form of effective field theory approach. To examine the interactions between weakly interacting massive particles(WIMPs) and Standard Model (SM) particles, we use the six-dimensional EFT mediated by new physics scale $\Lambda$ at tree level. After implementing a new effective field theory model in FeynRules \cite{Feynrules} We investigate the theory and constrain the theory by using relic density generated by MadDM\cite{Maddm} tool of MadGraph5\_aMC@NLO \cite{mg5}.}

\maketitle
\vfill

\section{Introduction}
\label{sec:intro}	

How to detect what happens when two particles interact has been well answered in modern physics. There is  one exception when an interaction includes dark matter (DM). At present we don't know what DM is and how it interacts with ordinary matter as well as there is not a sufficient theoretical model to answer these questions. On the other hand  we know some features of dark matter make us think that it is weakly interacting massive particle (WIMP)\cite{Massey}. 

If dark matter is particle, it must interact with ordinary matter weakly so we don't feel it electromagnetically. We are looking for a dead dark cat in a pitch-dark room hoping it exists; no light, no any kind of emission fortunately it has mass. Thereby observations about its existence rely on secondary implications from astrophysical and cosmological data. All of these cosmological and astrophysical implications can not be explained by other theoretical approach, but the presence of dark matter. Nevertheless the phrase of dark matter seems to be unfortunate because this particle is careless to any particle and thermodinamically frozen in galactic and cosmological scale. DM creation/annihilation process resume up to specific temperature meaning DM can not interact with something else and it becomes transparent electromagnetical sense. After that moment we can feel its hidden gravitational effect at the galactic and cosmic scale such as rotation curve of some galaxies imply the existence of it. Besides the amount of DM is significant to explain tensions over the value of Hubble parameter in the context of EFT in modified gravity  \cite{Cai1,Cai2,Cai3}.   

 According to the observations, \%24 of the universe is made up of this non-luminous and non-baryonic matter \cite{WMAP}.  Yet, there is poor information about its characteristics, such as its spin and momentum. There is no dark matter candidate in Standard Model. However, as it mentioned before, the dark matter candidate is supposed to be a weakly interacting massive particles which interact with SM particles. Beyond the SM, there are many models propose candidate of dark matter. One of the most popular candidate is the lightest stable supersymmetric particle of R-parity conserved supersymmetric model \cite{Jungman},\cite{Ellis},\cite{Silk}. Another popular candidates are, the stable lightest Kaluza-Klein photons of Kaluza-Klein parity conserved universial extra dimesion model \cite{Tait1}\cite{Tait2}, and the lightest T-odd particle in T-parity conserved little Higgs model \cite{Cheng},\cite{Birkedal}. Many other candidates has been proposed in the literature including scalar, fermion or vector DM mediators that can either belong to the SM, or can themselves belong to the hidden sector \cite{Mattelaer},\cite{Das}. In order to determine the link between dark sector and Standard model in a model independent approach, one can also describe an effective field theory. EFT consists of heavy mediator compressed with higher dimensional operators to describe the interactions between DM and SM particles\cite{Artz},\cite{Buchm}. EFT lagrangian with higher dimensional operators possess a new physics scale $\Lambda$, and validity of higher dimensional EFT basically depends on the relation between this new physics scale $\Lambda$ and transferred momentum $Q_{tr}$ of heavy mediator which for on-shell production of dark matter turns out to be $Q_{tr} \geq 2 m_D$. Recently, there are comprehensive studies on validity of EFT of dark matter \cite{Busoni} \cite{Bauer}. Apparently, minimal condition for new physics scale is required to be $\Lambda \geq 2 m_D$ \cite{Fermilab}. 

The most generic form of effective field theory lagrangian is formed by a set of interactions can possibly involve scalar, pseudoscalar, vector, axial vector and, tensor interactions.

\begin{equation}
\mathcal{L} \supset \sum_{i}^{} \mathcal{C}_i \mathcal{O}_i
\end{equation}

where $\mathcal{C}_i$ are set of coefficients in inverse mass dimension, and $\mathcal{O}_i$ are a set of operators formed by gauge-invariant combination of dark matter and SM fields. Dark matter field can be a scalar, fermion, or vector field, preserve quantum numbers of relevant symmetries. For simplicity, we will consider scalar and vector interactions between dark matter and SM particles.

In this work $SU(3)_c \times SU(2)_L \times U(1)_Y$ invariant six-dimensional operators in tree-level are focused on,dark matter field is considered to be a fermion field. We adopted on the operators in \cite{ref} to implement the model in FeynRules \cite{Feynrules}. The paper is organized as follows. In Sec. \ref{sec:model} the EFT operators are introduced and basic parameter of the theory are constrained  to provide the proper relic density. In Sec. \ref{sec:production} we present the result of monojet and dijet final states with missing $\slashed{E}_T$.  We compare that results of the productions for 13 TeV in Large Hadron Collider (LHC).

\section{Fermionic Dark Matter in Effective Field Theory and Relic Abundance}\label{sec:model}

\begin{figure}[H]
	\centering
	\includegraphics[scale=0.2]{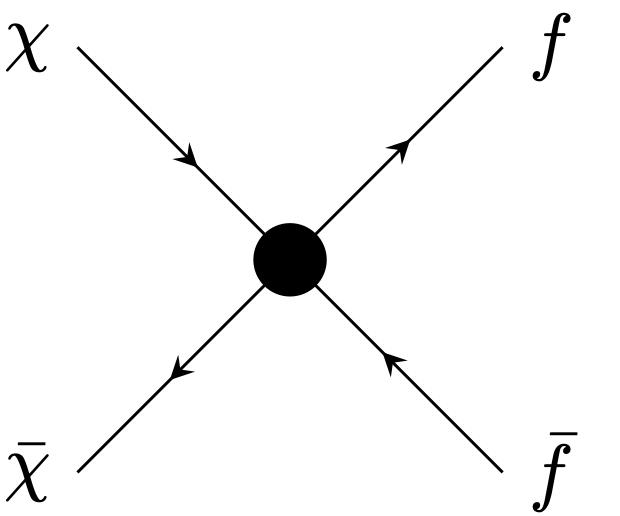}\hspace{0.5cm}
\includegraphics[scale=0.2]{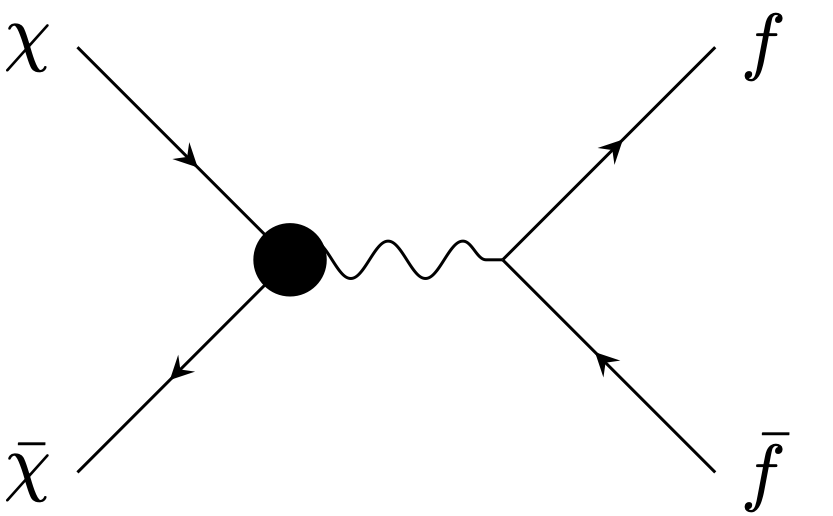}\hspace{0.5cm}
	\includegraphics[scale=0.2]{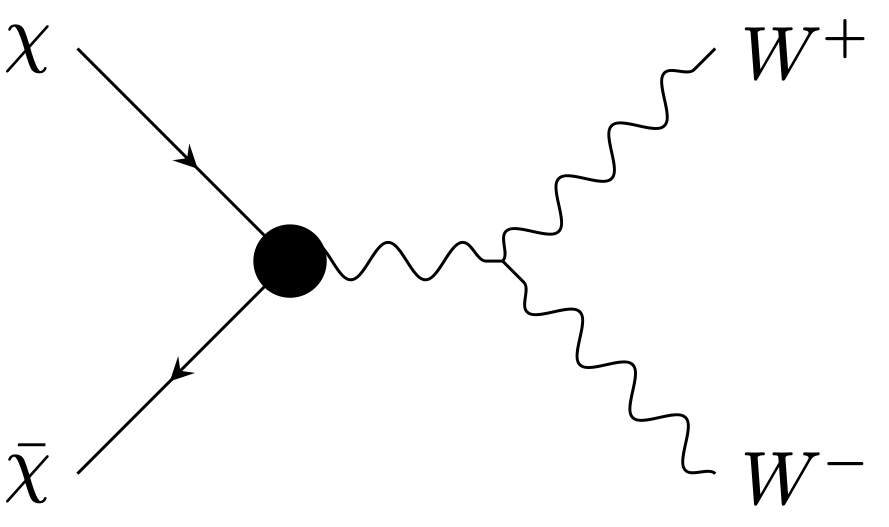}\hspace{0.5cm}
	\includegraphics[scale=0.2]{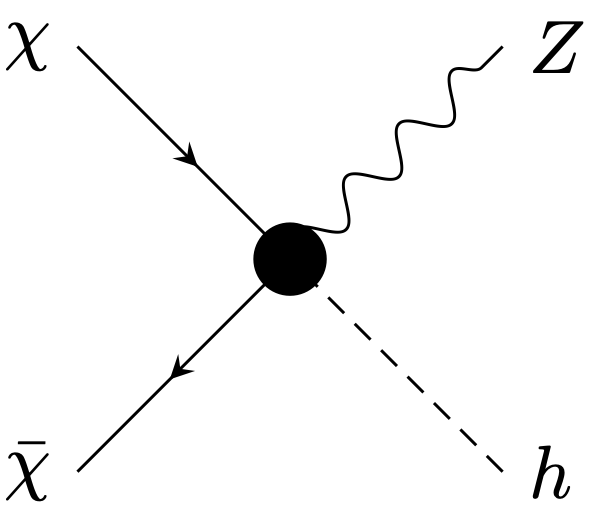}\hspace{0.5cm}
	\includegraphics[scale=0.2]{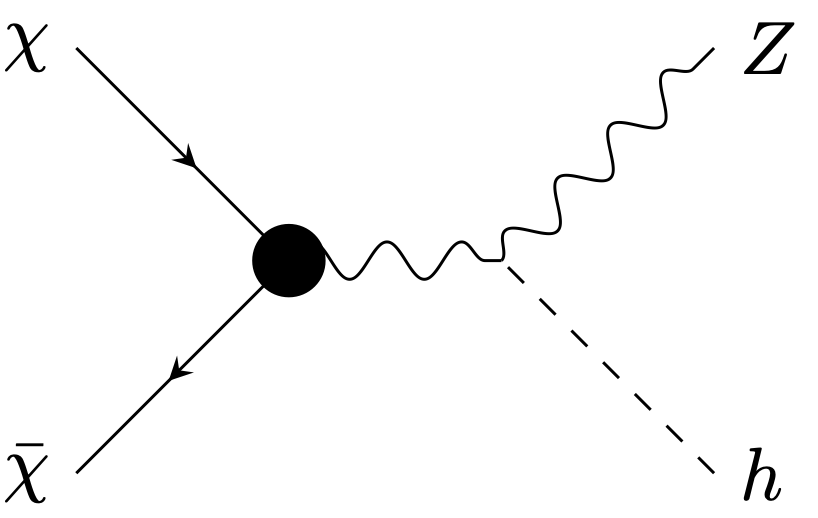}\hspace{0.5cm}
	\caption{Feynman diagrams for fermionic dark matter annihilation of six-dimensional EFT. Effective vertices are indicated by dots.}
	\label{fig:feynman}
	\end{figure}

In this section, fermionic dark matter is described in six-dimensional effective field theory (EFT), briefly. To preserve the validity of EFT, the cut-off scale $\Lambda$ is constrained to be at least two times bigger than the mass of dark matter candidate due to the momentum transfer \cite{Fermilab}. It is assumed that the dark matter is a  SM gauge singlet Dirac fermion, odd under $Z_2$ symmetry, and, for interactions between SM and DM, tree level $SU(3)_c \times SU(2)_L \times U(1)_Y $ invariant dimension-six operators of ref. \cite{ref} are used. The main annihilation processes for fermionic dark matter in EFT are depicted in Fig. \ref{fig:feynman}. The filled circles denote effective vertices replaced with heavy mediator in the theory where Feynman diagrams are based on the operators are manifested in ref. \cite{ref}. The first and the second diagrams,respectively, show the four-fermion effective interactions and fermion interactions via vector mediator which interacts fermionic dark matter, effectively. And, the other diagrams demonstrate the electroweak interactions of fermionic dark matter in EFT.

	\subsection{Method for Analysis of DM Candidates}

DM researches are simply based on four parameters in EFT scope: relic density, coupling constant, the mass of DM and the scale of physics. Among them the relic density is the key; experimentally measured and reliable, so it drives the discussions about the true nature of this unknown matter. Naturally we follow the outline drawn by taking the relic density as an anchor in order to retain the most convenient values of other parameters. Thanks to very useful tools which have been already developed we embed six-dimensional operators (now accessible to all higher dimensional operators) into the related tools , described below, to discuss the values of these arbitrary parameters as show up in the lagrangian.  

6-dimensional effective field theory model is probed by scanning the fundamental parameters, such as, mass of dark matter candidate, coupling parameters and cut-off scale. We use  \textsc{MadDM} package \cite{Maddm} to compute the relic density after introducing the model in \textsc{FeynRules} \cite{Feynrules}. We defined  field of the dark matter, and all its characteristics, all of the couplings, kinematical constants and interaction Lagrangians into FeynRules. After applying hermicity checks and calculating Feynman rules. We eventually managed to obtain the UFO model output to calculate relic density and scan parameters in \textsc{MadDM}. 
4-fermion vectoral Lagrangian can be explicitly represented as:

		\begin{eqnarray}
		\mathcal{L}_{(u_R,d_R,e_R)\chi}=\frac{g^u_{R}}{2 \Lambda^2}(\bar{u}\gamma^{\mu}u)(\bar{\chi}\gamma_{\mu}\chi) +\frac{g^d_{R}}{2 \Lambda^2}(\bar{d}\gamma^{\mu}d)(\bar{\chi}\gamma_{\mu}\chi) + \frac{g^e_{R}}{2\Lambda^2}(\bar{e}\gamma^{\mu}e)(\bar{\chi}\gamma_{\mu}\chi)  
		\end{eqnarray}
	
		where $\Lambda$ is effective field theory cut-off scale $u_R$, $d_R$ and $e_R$'s represent right-handed fermions and, $g_R$'s are the coupling constants of each type of fermion-singlet with fermionic dark matter, given Appendix A.
		
4-fermion scalar Lagrangian can be represented as:

		\begin{eqnarray}
		\mathcal{L}_{(\ell,q)\chi}=\frac{g^{\ell}_L}{\Lambda^2}(\bar{\ell}\chi)(\bar{\chi}\ell)  + \frac{g^q_L}{\Lambda^2} (\bar{q}\chi)(\bar{\chi}q)
		\end{eqnarray}
		
		where $\ell$ denotes lepton-doublet and $q$ denotes left-handed quark-doublet/ $g^{\ell}_L$ and $g^{q}_L$ are coupling constants are leptons and quarks with fermionic dark matter.

Vector, scalar and fermion interaction Lagrangian can take the form: 

\begin{eqnarray}
		\mathcal{L}_{\phi\chi}=\frac{i \alpha_{\phi \chi}}{\Lambda^2}({\phi}^{\dagger}D^{\mu}\phi)(\bar{\chi}\gamma_{\mu}\chi)+h.c.
		\end{eqnarray}

where, $\Lambda$ is scale of new physics\ $\phi$ is Higgs doublet and, $g_{R}^{u(d,e)}$'s and $g_L^{\ell(q)}$ are coupling parameters related to the dark operators $\alpha$'s, for each type of fermion, are  given in ref. \cite{ref} and  Appendix A.


Because the relic density is the most crucial quantity to investigate any dark mater model, in six-dimensional EFT of fermionic dark matter model, we examine the behaviours of $\alpha$ model parameters, $\Lambda$ cut-off scale and, mass of fermionic dark matter $M_{\chi f}$ with respect to the relic density. In this work, parameters are scanned individually, and are constrained to approach proper relic density.

In the first part of our analysis we focus on mass of fermionic dark matter candidate. We set the $\alpha$ parameters as $\alpha_{u \chi}=\alpha_{d \chi}=\alpha_{e \chi}=\alpha_{q \chi}=\alpha_{l \chi}=\alpha_{\phi \chi}=1$ and scan the mass up to $1$ TeV.

\begin{figure}[H]
	\centering
	\includegraphics[scale=0.70]{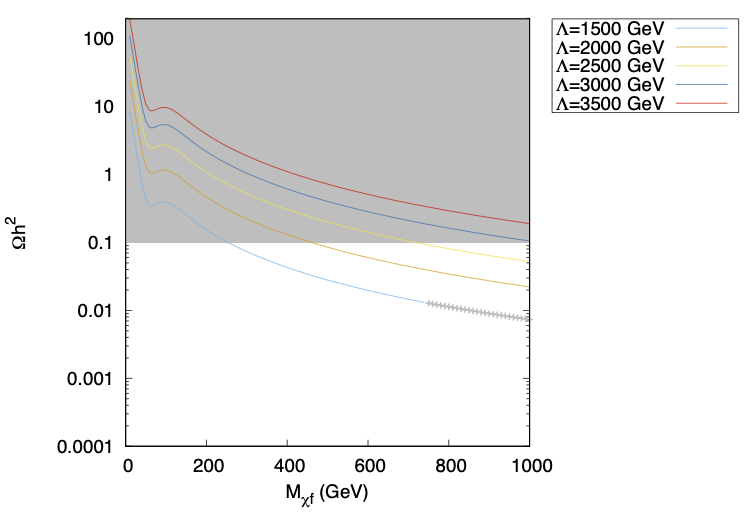}
	\includegraphics[scale=0.70]{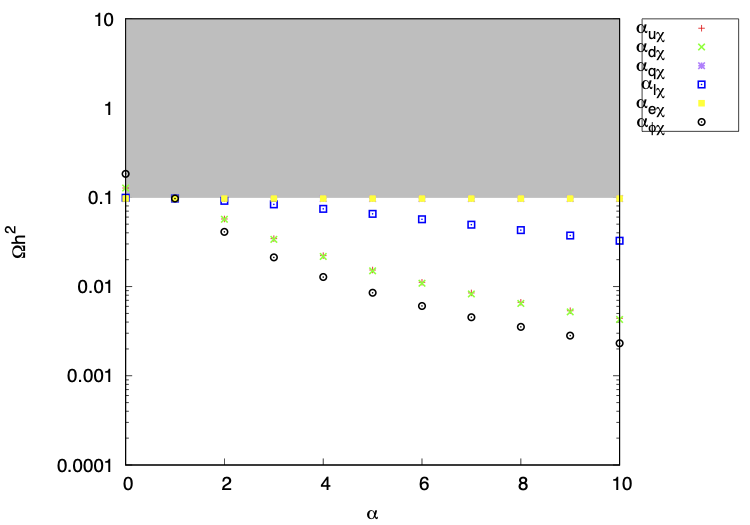}\
	\caption{a) Relic density $\Omega h^2$ vs. mass of fermionic dark matter candidate $M_{\chi f}$. b) Model parameters $\alpha$'s vs. relic density $\Omega h^2$. Shaded region denotes the excluded region.}
	\label{fig:mxfrel}
\end{figure}

Secondly, we adopted on the $\alpha$ parameters in the analysis. In Fig. \ref{fig:mxfrel}, we demonstrated the effect of each $\alpha$ parameter, individually. Mass of fermionic dark matter candidate $M_{\chi f}$ is taken $460$ GeV and, new physics scale $\Lambda$ is set to $2000$ GeV.

\begin{figure}[H]
\centering
	\includegraphics[scale=0.65]{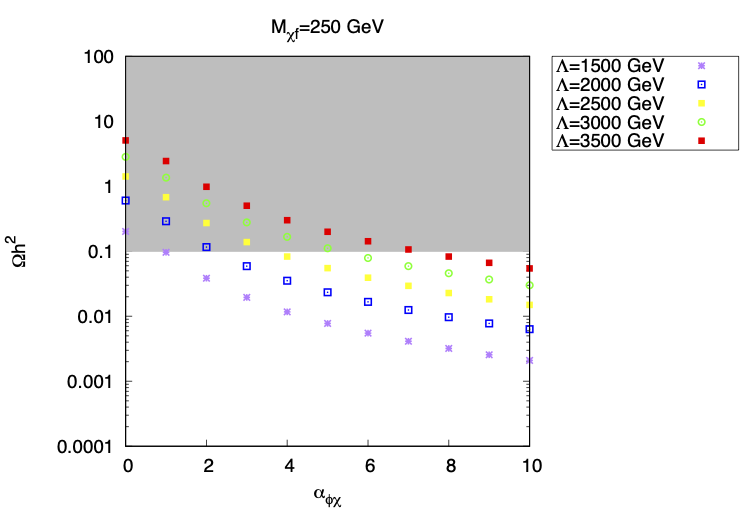}
 \includegraphics[scale=0.65]{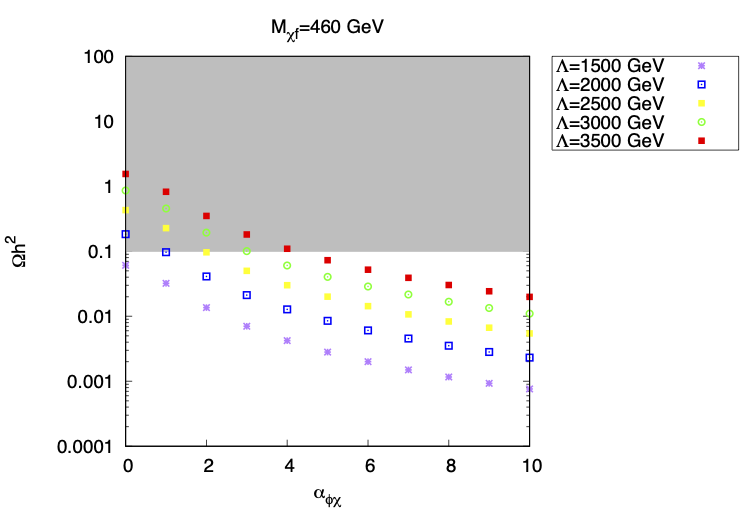}
	\includegraphics[scale=0.65]{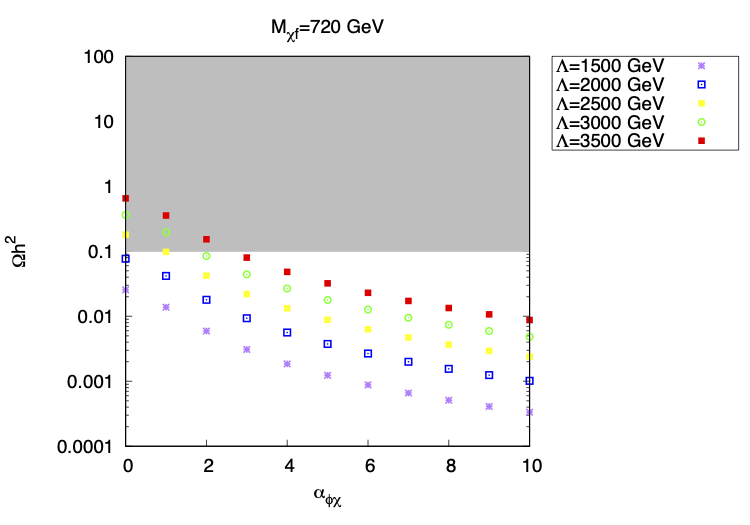}\
	\caption{$\alpha_{\phi \chi}$'s vs. relic density $\Omega h^2$. Shaded grey region denotes the model exluded region.}
	\label{fig:alphaphirel}
	\end{figure}

 In Fig. \ref{fig:mxfrel} when different values of $\Lambda$ is taken into account,the behaviour of the relic density versus $M_{\chi f}$ does not indicate any change. The shaded region in Fig. \ref{fig:mxfrel} demonsrates the excluded region of six-dimensional EFT. And the notched gray line demonstrates excluded region as a consequence of requied condition,$\Lambda \geq2 m_D$.  One can notice that, as $\Lambda$ grows, model lands to the excluded region. And, as the mass of dark matter increases, the model is prone to be out of the excluded region.  It is also obvious from the Fig.\ref{fig:mxfrel} that $\alpha_{\phi \chi}$ is the most dominant parameter on account of the electroweak interaction of fermionic dark matter in EFT. It can also be inferred from the Fig. \ref{fig:mxfrel} that right-handed electrons and left-handed quarks do not give any contributions to relic density. 
The shaded region in the Fig. \ref{fig:mxfrel} (b) demonstrates excluded region of the six-dimensional fermionic EFT. One can easily notice that the biggest contribution to relic density comes from fermion-vector-scalar interaction. Conversely, $\alpha_{u\chi}$ and $\alpha_{q \chi}$ does not give any contribution to relic density. Because the most considerable contribution to relic density comes from the $\alpha_{\phi\chi}$, we probe how relic density changes due to $\alpha_{\phi \chi}$ parameter. In Fig \ref{fig:alphaphirel} we scan $\alpha_{\phi \chi}$ for some specific values of mass $M_{\chi f}$ and $\Lambda$, and, set all other $\alpha$ parameters to $1$. 

As it can be seen in Fig. \ref{fig:alphaphirel} the behaviour of relic density with respect to $\alpha_{\phi \chi}$ does not change. However, as the mass $M_{\chi_f}$ and $\Lambda$ increase, the model goes to relic density exclusion limits. As the mass increases the relic density shifts to appropriate regime for relic density, converserly, as cut-off increases it turns out to be unreasonable limits for relic density.

We also examined the relic density of model against  mass of dark matter $M_{\chi f} $, cut-off scale $\Lambda$ and model parameter $\alpha_{\phi \chi}$, simultaneously. $M_{\chi f}$ and $\alpha_{\phi \chi}$ are scanned with specific values of $\Lambda=1500$ GeV $\Lambda=2000$ GeV and $\Lambda=2500$ GeV.

\begin{figure}[H]
	\centering
	\includegraphics[scale=0.5]{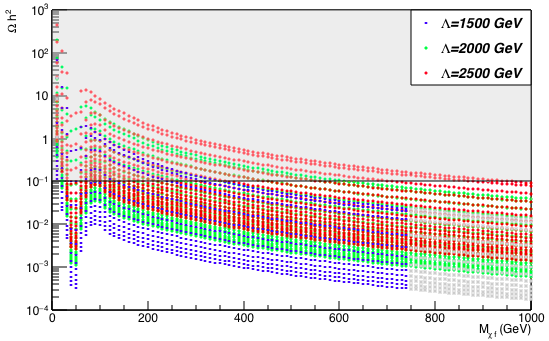}
	\label{fig:DMscanned}
		\caption{$\alpha_{\phi \chi}=0.2,......,10$ is scanned. Gray data points and gray-shaded region indicates the excluded region according to current observed relic density.}
\end{figure}

In Fig.  \ref{fig:DMscanned} $M_{\chi f}$ blue, green and red dots, eacf for $\Lambda=1500$ GeV $\Lambda=2000$ GeV and $\Lambda=2500$ GeV. respectively, give relic density each by differing $M_{\chi} f$ and $\alpha_{\phi \chi}$.  As cut-off scale  $\Lambda$ increases, model intensely affected and relic density value increases, too.

\section{Dark Matter Production}\label{sec:production}

\begin{figure}[h!]
\centering
\includegraphics[scale=0.3]{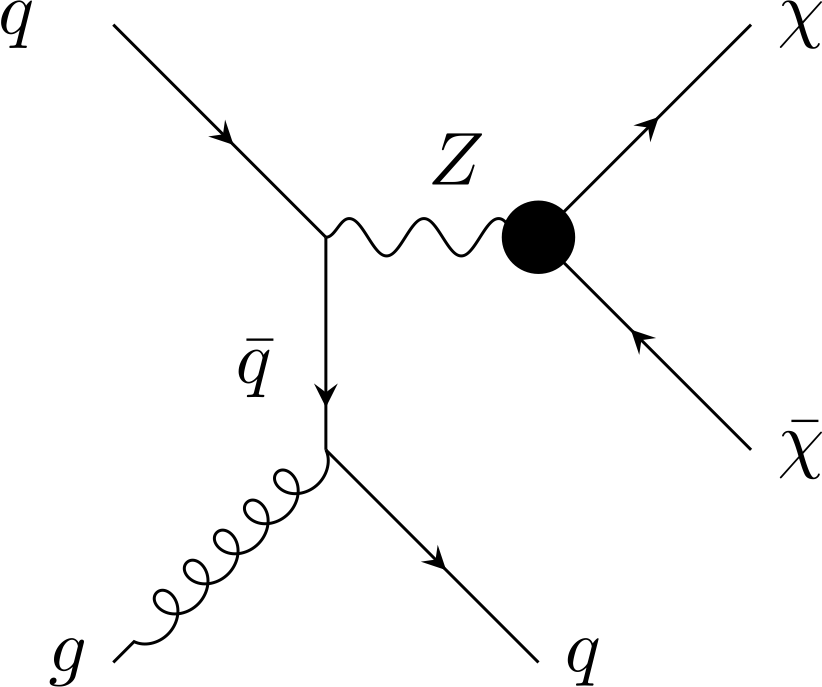} \qquad \qquad \qquad
\includegraphics[scale=0.3]{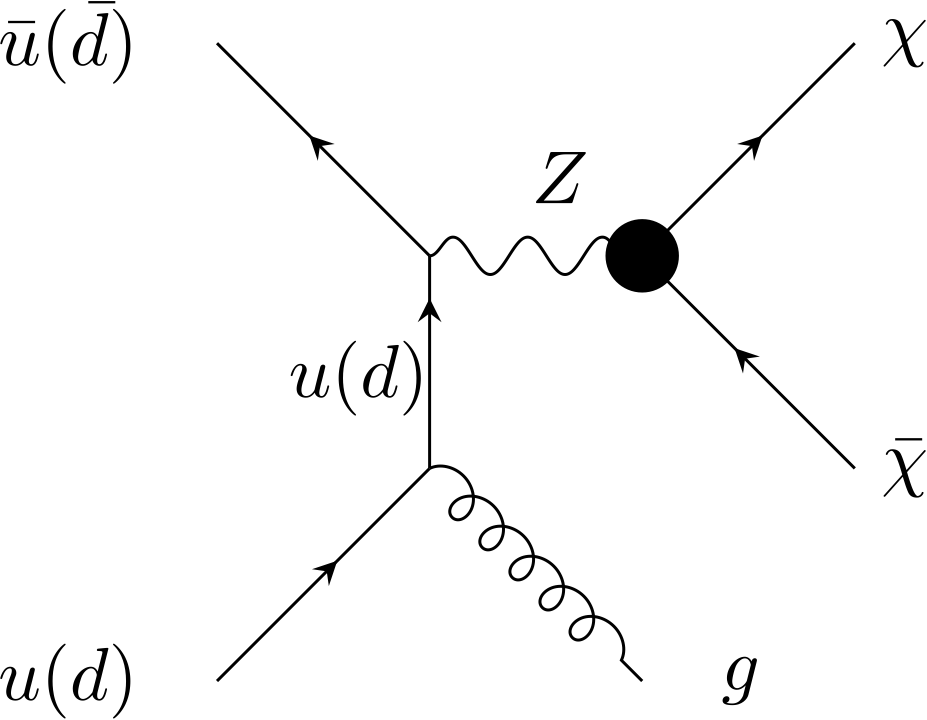}
\caption{Feynman diagrams of the most dominant processes which contribute to $pp \rightarrow \bar{\chi} \chi + j$}
\label{fig:monojet-feynman}
\end{figure}

For dark matter pair-production, commonly jets accompany to missing transverse energy. Monojet and dijet with missing transverve energy final states are the most dominant processes for pair production of dark matter. In this work, for different cut-off values of $\Lambda$  the process $\chi \bar{\chi} j$ and  $\chi \bar{\chi} j j$ final states are analysed. Jet $p_T$'s, jet $\eta$'s and missing $E_T$'s are obtained using  \textsc{MadGraph5\_aMC@NLO}.  Dark matter pair-production processes with $\slashed{E}_T+monojet$ final states are shown in Fig. \ref{fig:monojet-feynman}

The cross sections of $\slashed{E}_T+j$ final state for $13$ TeV center of mass energy in LHC, are listed in the Table \ref{tab:monojet}.

\begin{table}[H]
\scalebox{0.8}{
\centering
\begin{tabular}{|c|c|c|c|}
\hline 
$M_{\chi f}$ & $\sigma_{pp>\chi \bar{\chi} j}(pb),$ $\Lambda=1500$ GeV & $\sigma_{pp>\chi \bar{\chi} j}(pb),$ $\Lambda=2000$ GeV & $\sigma_{pp>\chi \bar{\chi} j}(pb),$ $\Lambda=2500$ GeV \tabularnewline
\hline 
\hline 
$M_{\chi f}=250$ GeV & $2.76$ $10^{-2}$ & - & -\tabularnewline
\hline 
$M_{\chi f}=460$ GeV  &  $1.67$ $10^{-2}$&  $5.31$ $10^{-3}$ & -\tabularnewline
\hline 
$M_{\chi f}=720$ GeV & $8.75$ $10^{-3}$ &  $2.76$ $10^{-3}$ &  $1.13$ $10^{-3}$\tabularnewline
\hline 

\end{tabular}
}
\caption{Cross section list of $p\ p > \bar{\chi} \chi \ j$ process, for 13 TeV center of mass energy of LHC.}
\label{tab:monojet}
\end{table}

\begin{figure}[H]
\centering
\includegraphics[scale=0.30]{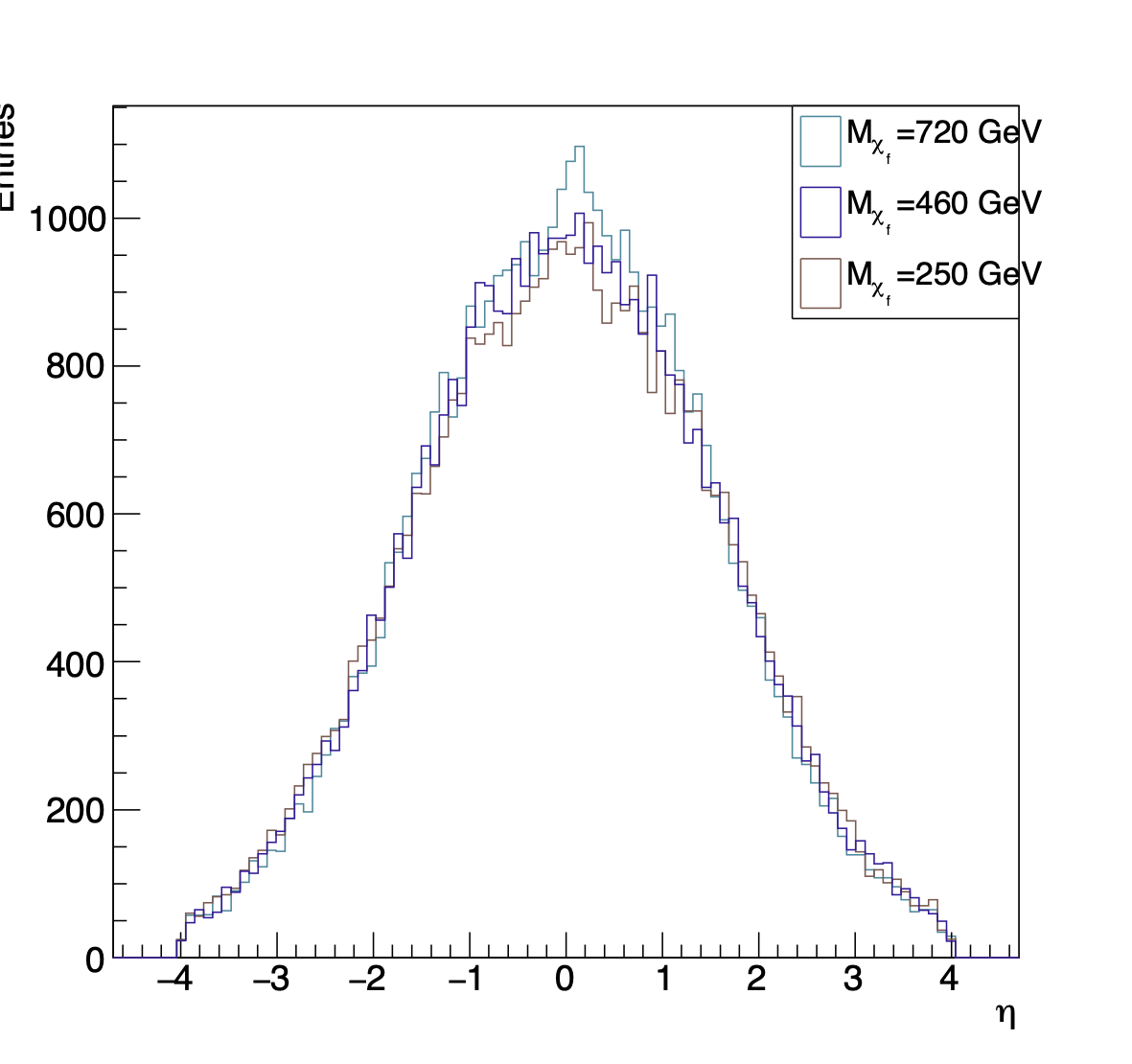}\
\includegraphics[scale=0.35]{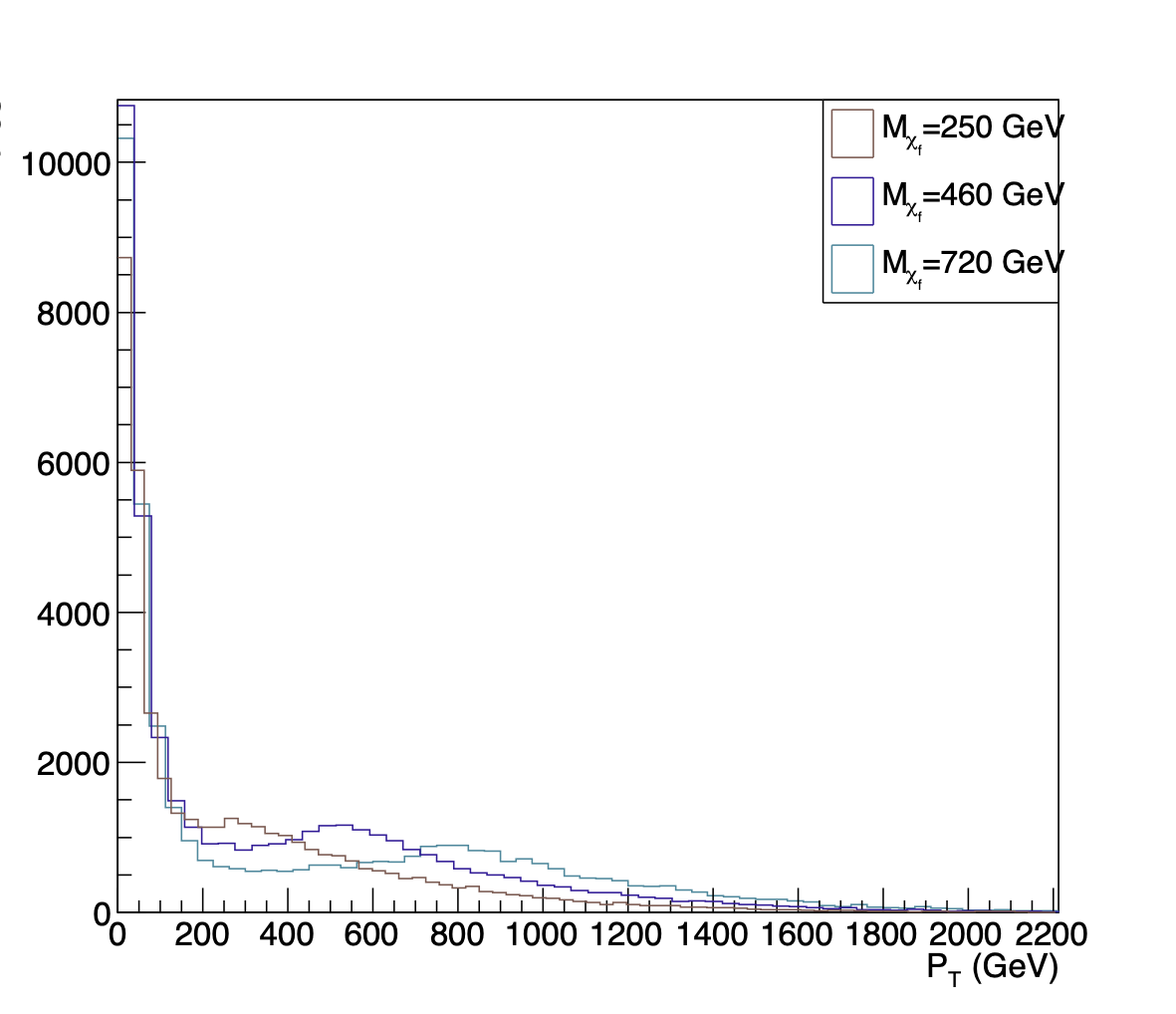}\\
\includegraphics[scale=0.40]{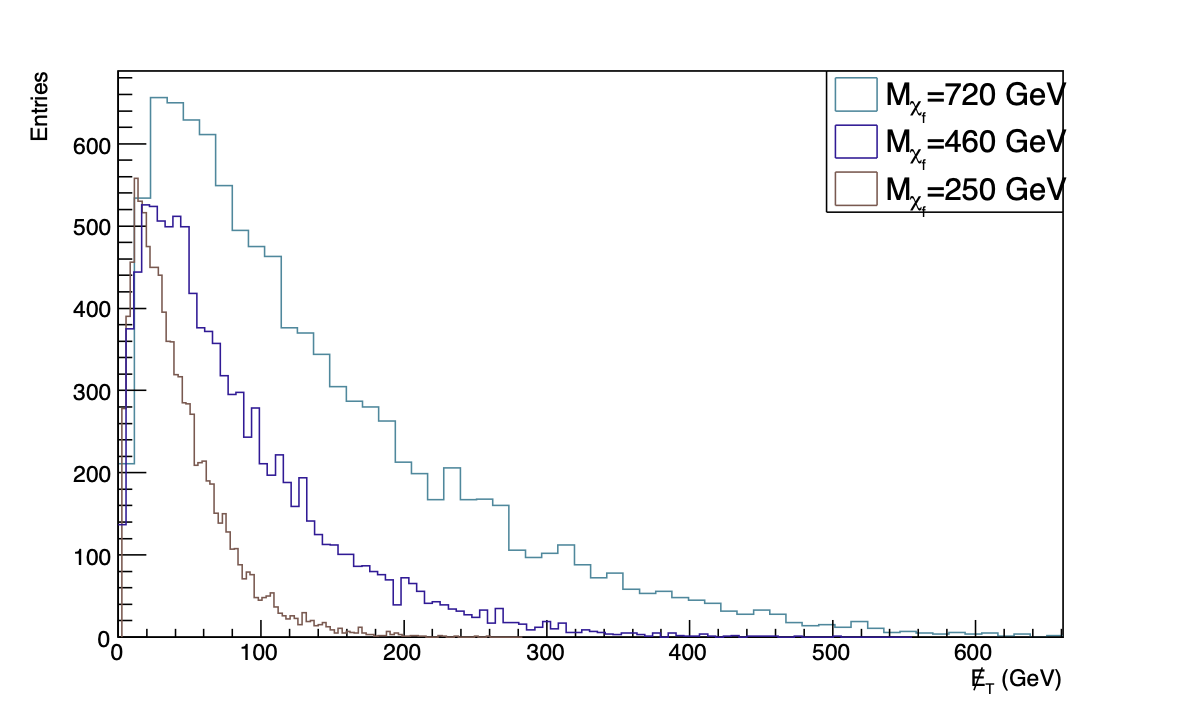}
	\caption{$p_T$, $\eta$ and $\slashed{E}_T$ for $p p \rightarrow \chi \bar{\chi} j $ for  $M_{\chi_f}= 250$ GeV, $M_{\chi_f}= 460$ GeV, and $M_{\chi_f}= 720$ GeV when $\Lambda=1500$ GeV.}
	\label{fig:monojet_lambda1500}
	\end{figure}

Dark matter pair production with monojet, for 13 TeV center of mass energy of LHC are obtained for 10000 events with standard cuts of energy. Consequently, in Table \ref{tab:monojet} it can be figured out that while the mass and cut-off increase, production cross section decreases for pair production of dark matter. For $\Lambda=1500$ GeV and $\alpha_{\phi \chi (u\chi, d\chi, e \chi, q \chi, \ell \chi)}=1$,  $M_{\chi_f}\geq 250$  GeV values provide the proper relic density. The  $\eta$ and $p_T$'s of jet and missing transverse energy of dark matter for pair production process are result in Fig. \ref{fig:monojet_lambda1500}.

From Fig. \ref{fig:monojet_lambda1500} it can be noted that the $\eta$'s and $p_T$'s do not chance according to the mass of the dark matter candidate. This is obvious result of final state jet, which is not originated from dark matter mediator. However, missing transverse energy changes to the mass of dark matter candidate, since it is directly related to the mass of dark matter. For the values of $\Lambda=2000$ GeV and $\alpha_{\phi \chi (u\chi, d\chi, e \chi, q \chi, \ell \chi)}=1$, when $M_{\chi_f}\geq 460$ GeV relic density comes up to reasonable regime. 

\begin{figure}[H]
\centering
\includegraphics[scale=0.40]{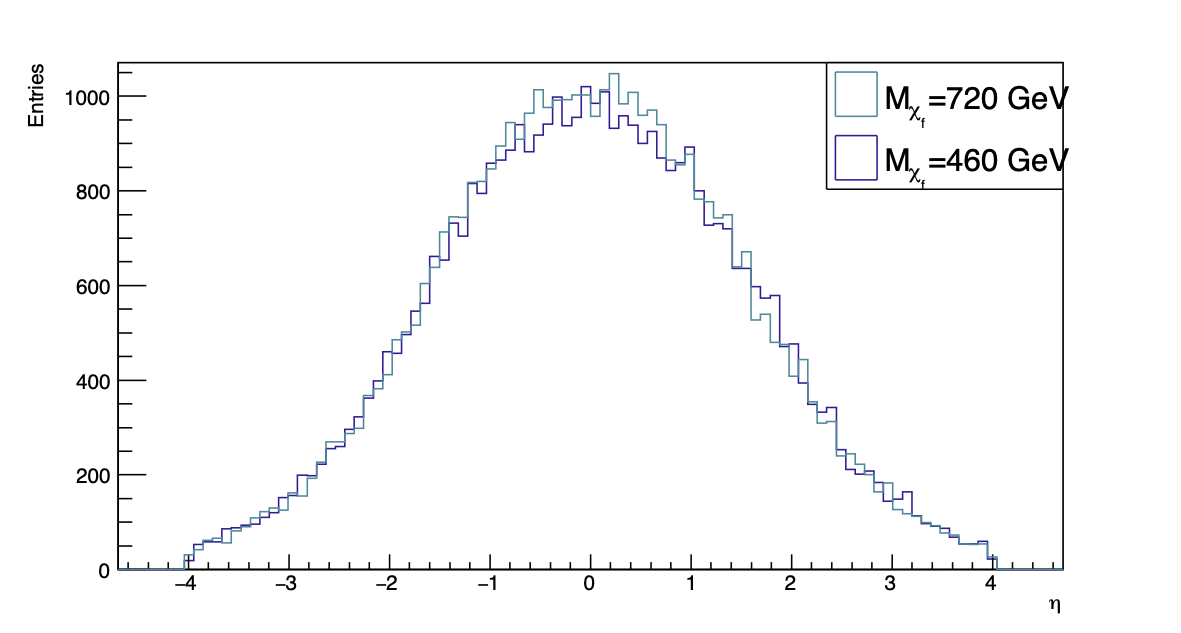}\
\includegraphics[scale=0.40]{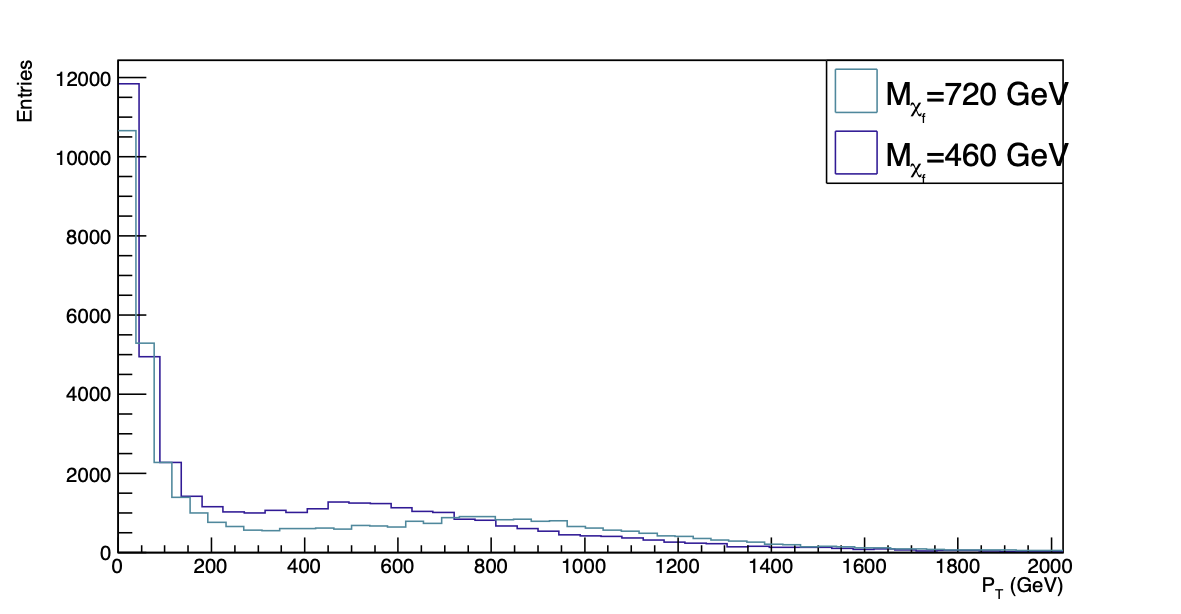}\
\includegraphics[scale=0.45]{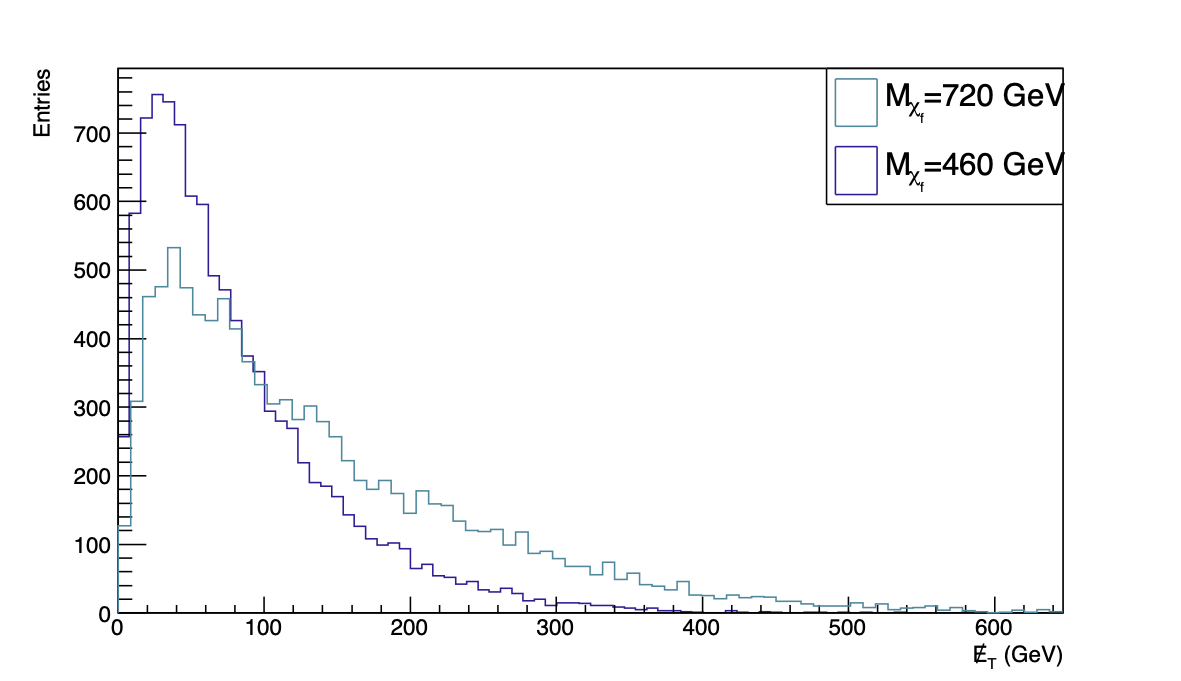}
\caption{$p_T$, $\eta$ and $\slashed{E}_T$ for $p p \rightarrow \chi \bar{\chi} j $ for  $M_{\chi_f}= 460$ GeV, and $M_{\chi_f}= 720$ GeV when $\Lambda=2000$ GeV.}
	\label{fig:monojet_lambda2000}
	\end{figure}

Since the final state jet is not originated from dark matter, in Fig. \ref{fig:monojet_lambda2000} $\eta$ and $p_T$ of jet do not change with respect to the mass of dark matter. But missing $\slashed{E}_T$ changes according to the mass of dark matter, which is based on dark matter itself.

When one takes the $\Lambda=2500$ GeV and $\alpha_{\phi \chi (u\chi, d\chi, e \chi, q \chi, \ell \chi)}=1$, $M_{\chi_f}\geq 720$ GeV values give the proper relic density. 

\begin{figure}[H]
\centering
\includegraphics[scale=0.30]{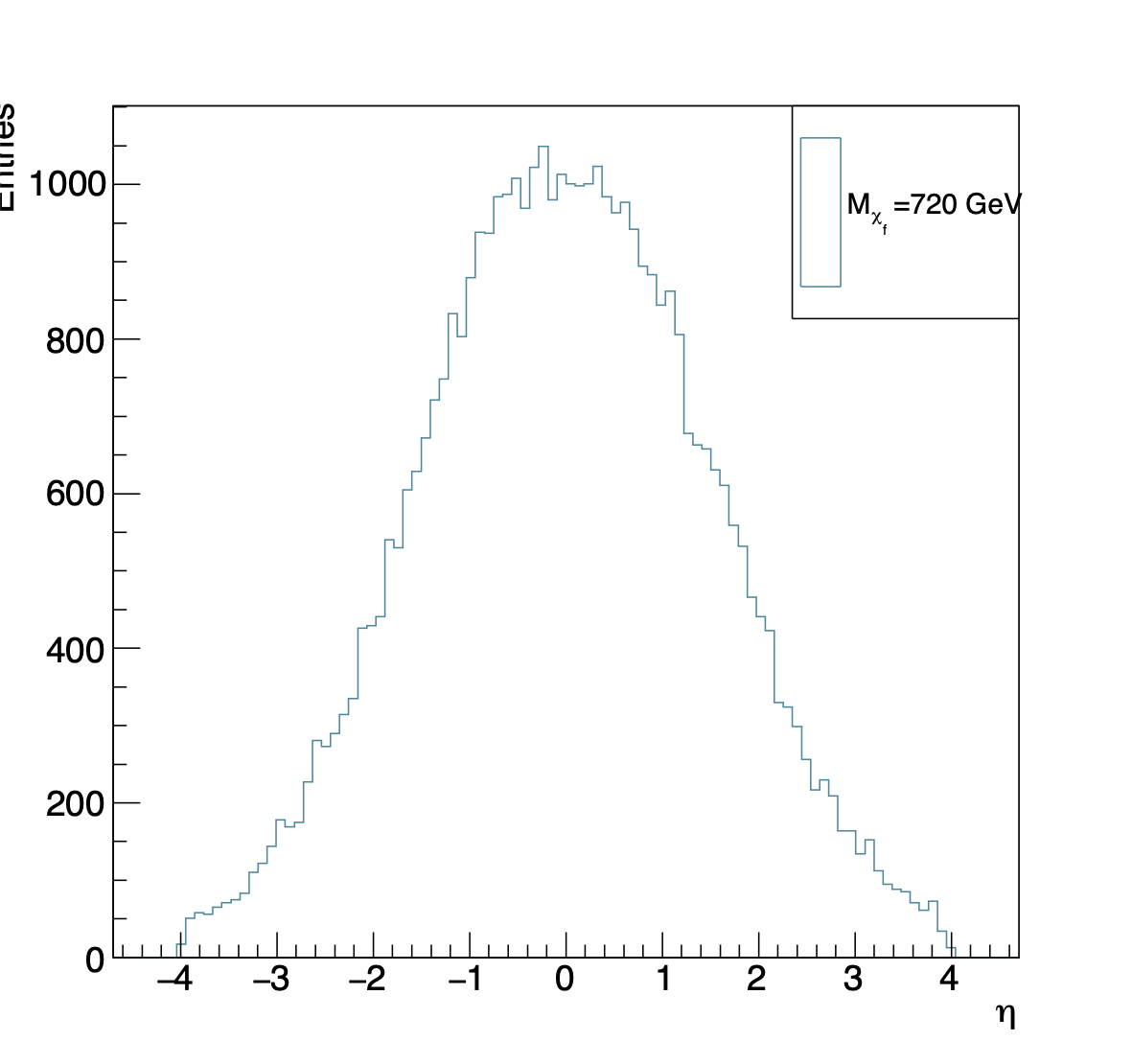}\
\includegraphics[scale=0.45]{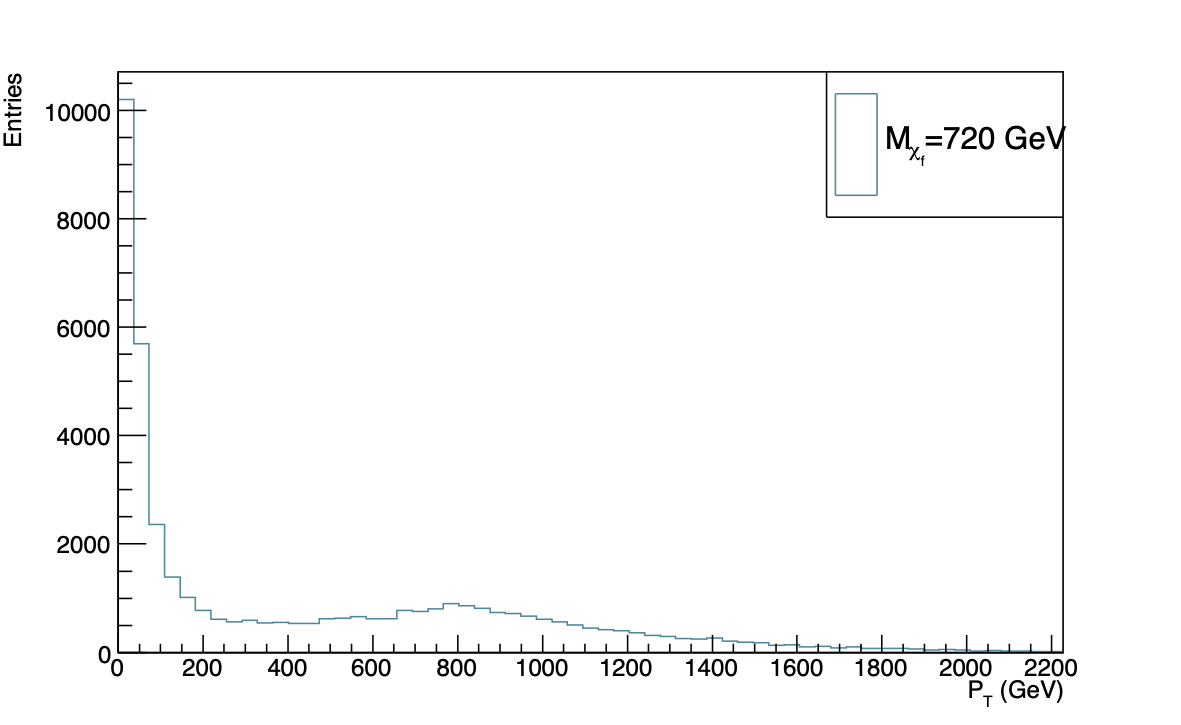}\
\includegraphics[scale=0.45]{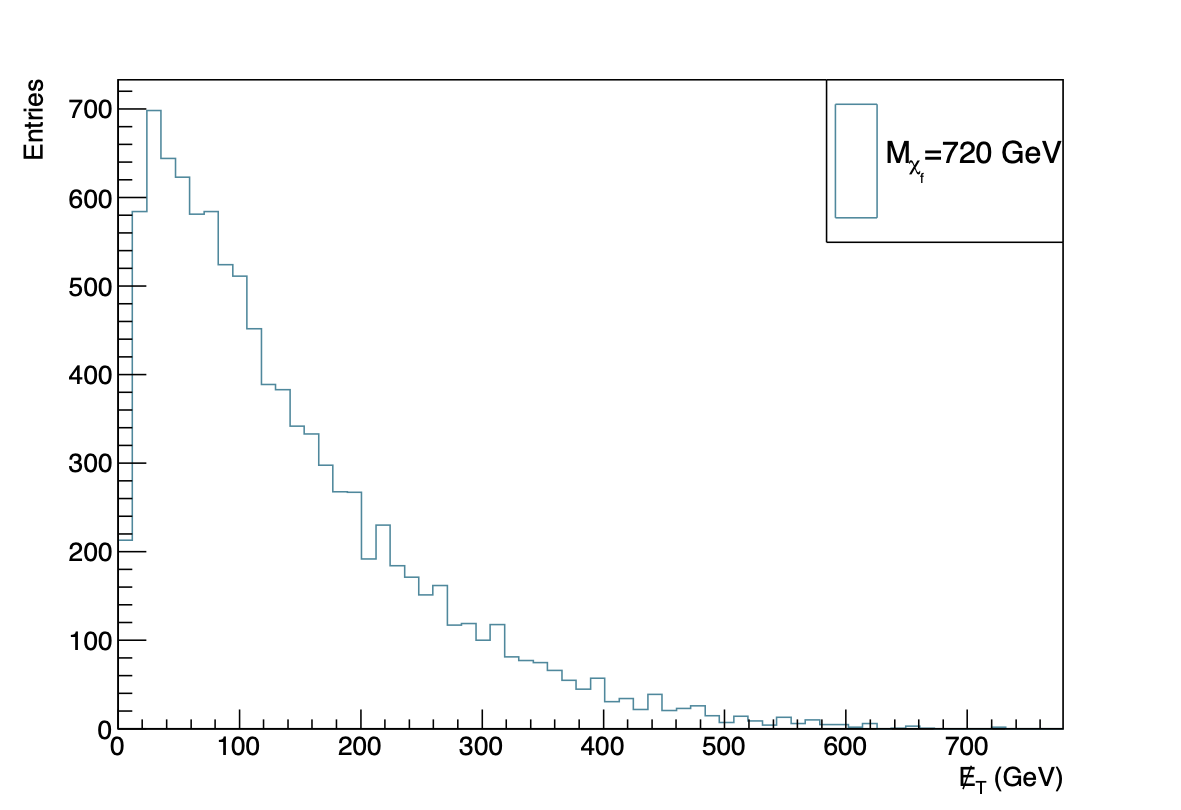}
	\caption{$p_T$, $\eta$ and $\slashed{E}_T$ for $p p \rightarrow \chi \bar{\chi} j $ for  $M_{\chi_f}= 720$  GeV when $\Lambda=2500$ GeV.}
	\label{fig:monojet_lambda2500}
	\end{figure}
	
	Fig. \ref{fig:monojet_lambda2500} demonstrates the characteristics of $p_T$, $\eta$ and $\slashed{E}_T$ for monojet accompanying final states to dark matter pair production. $\slashed{E}_T$ takes lower value of when it compared with $\Lambda=1500$ GeV and $\Lambda=2000$ GeV.

\begin{figure}[H]
\centering
\includegraphics[scale=0.50]{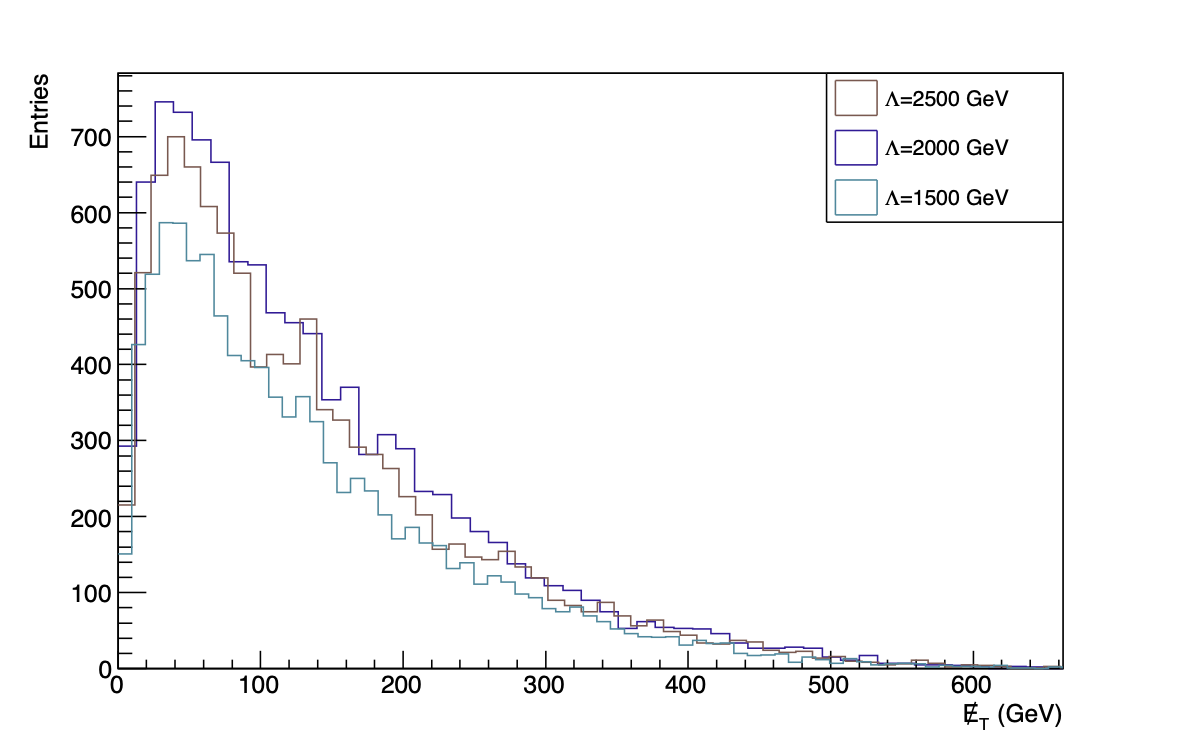} 
	\caption{$\slashed{E}_T$ for $M_{\chi}f=720$  GeV for $pp\rightarrow \bar{\chi} \chi j  $ for $13$ TeV center of mass energy of LHC.}
	\label{fig:monojet_mxf720}
	\end{figure}

Finally, cecause $M_{\chi_f}=720$ GeV provides the reasonable relic density for these $\Lambda$ values of, $\Lambda=1500 GeV$ $\Lambda=2000$ GeV and $\Lambda=2500$ GeV, and for  $\alpha_{\phi \chi (u\chi, d\chi, e \chi, q \chi, \ell \chi)}=1$, we compared the missing transverse energy of monojet process for these  $\Lambda$ values Fig. \ref{fig:monojet_mxf720}. 

\begin{figure}[H]
\centering
\includegraphics[scale=0.25]{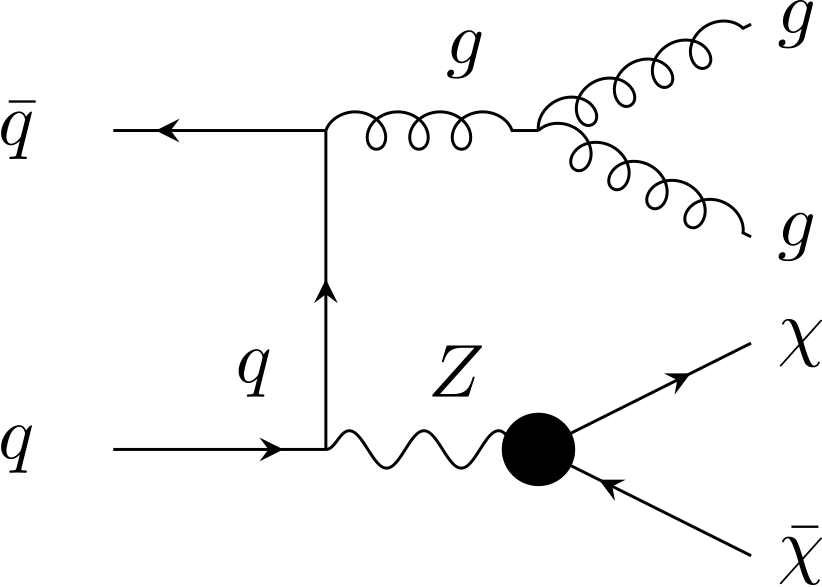} \qquad 
\includegraphics[scale=0.25]{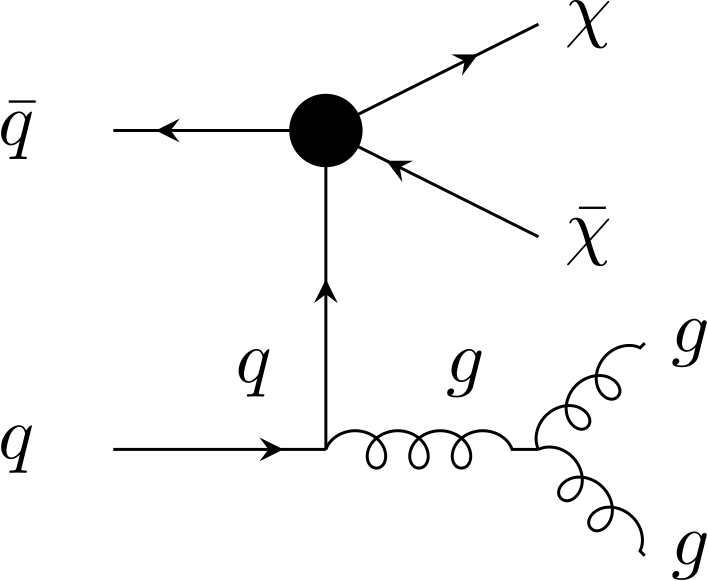} \qquad 
\includegraphics[scale=0.25]{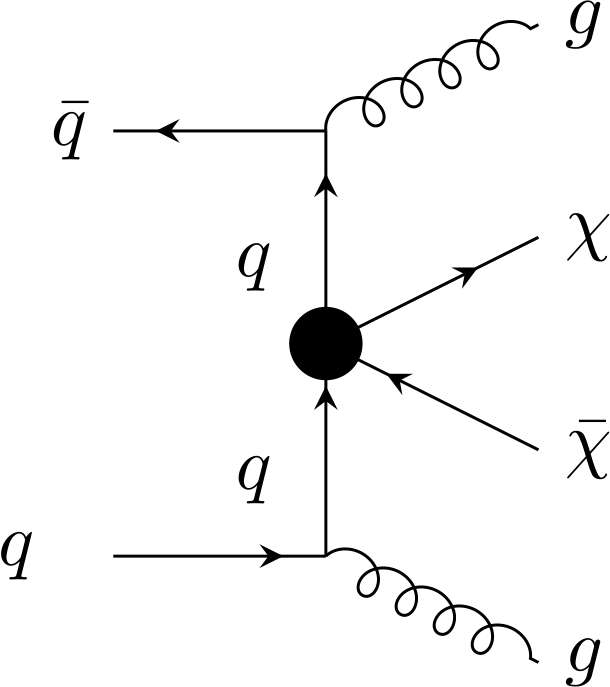} \qquad 
\includegraphics[scale=0.25]{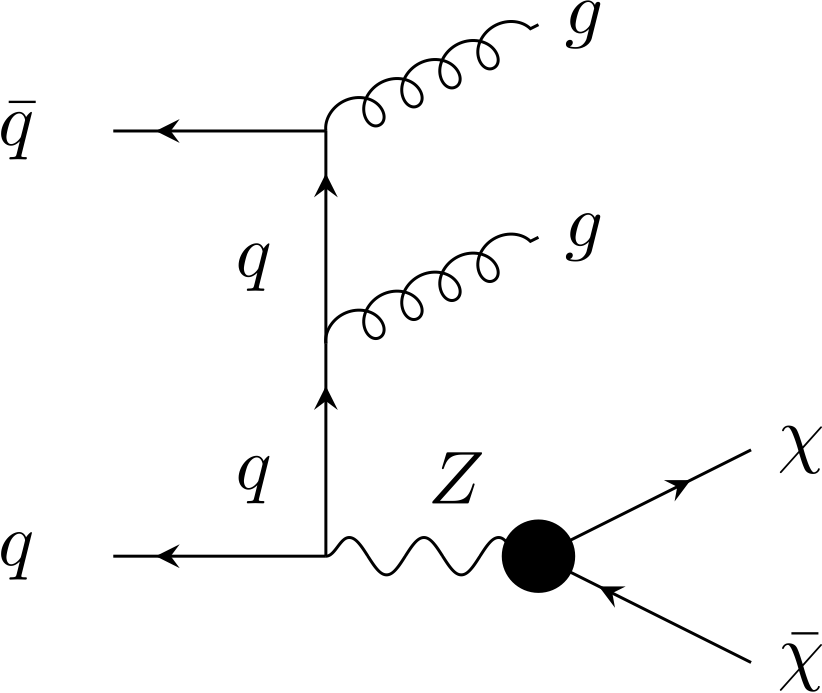} 
\label{fig:dijet-feynman}
\caption{Feynman diagrams of the most dominant processes which contribute to $pp \rightarrow \bar{\chi} \chi + jj$ process}
\end{figure}

In the second part of the work, we present a preanalysis for the possibility of the dark matter production via dijet process at recent colliders. Feynman diagrams which contribute dominantly to $\slashed{E}_T+jj$ processes are shown in Fig. \ref{fig:dijet-feynman}.

The cross sections of $\slashed{E}_T+jj$ final state for $13$ TeV center of mass energy in LHC, are listed in the Table \ref{tab:dijet}.

\begin{table}[H]

\scalebox{0.8}{
\begin{tabular}{|c|c|c|c|}
\hline 
$M_{\chi f}$ & $\sigma_{pp>\chi \bar{\chi} j j}(pb),$ $\Lambda=1500$ GeV & $\sigma_{pp>\chi \bar{\chi} j}(pb),$ $\Lambda=2000$ GeV & $\sigma_{pp>\chi \bar{\chi} j}(pb),$ $\Lambda=2500$ GeV \tabularnewline
\hline 
\hline 
$M_{\chi f}=250 GeV$ &  $1.33$ $10^{-2}$& - & -\tabularnewline
\hline 
$M_{\chi f}=460 GeV$  &  $8.28$ $10^{-3}$& $2.62$ $10^{-3}$ & -\tabularnewline
\hline 
$M_{\chi f}=720 GeV$ & $4.37$ $10^{-3}$&  $1.38$ $10^{-3}$ &  $5.65$ $10^{-4}$\tabularnewline
\hline 

\end{tabular}
}
\caption{Cross section list of $p\ p > \bar{\chi} \chi \ j \j$ process, for 13 TeV center of mass energy of LHC.}
\label{tab:dijet}
\end{table}

From Table \ref{tab:dijet} it can be noted that, for pair production of dark matter, while the mass and cut-off increase, production cross section decreases. 

Because none of the final state jets emerge from dark matter mediator, we focuces on the missing transverse energy $\slashed{E}_T$.

\begin{figure}[H]
\centering
\includegraphics[scale=0.40]{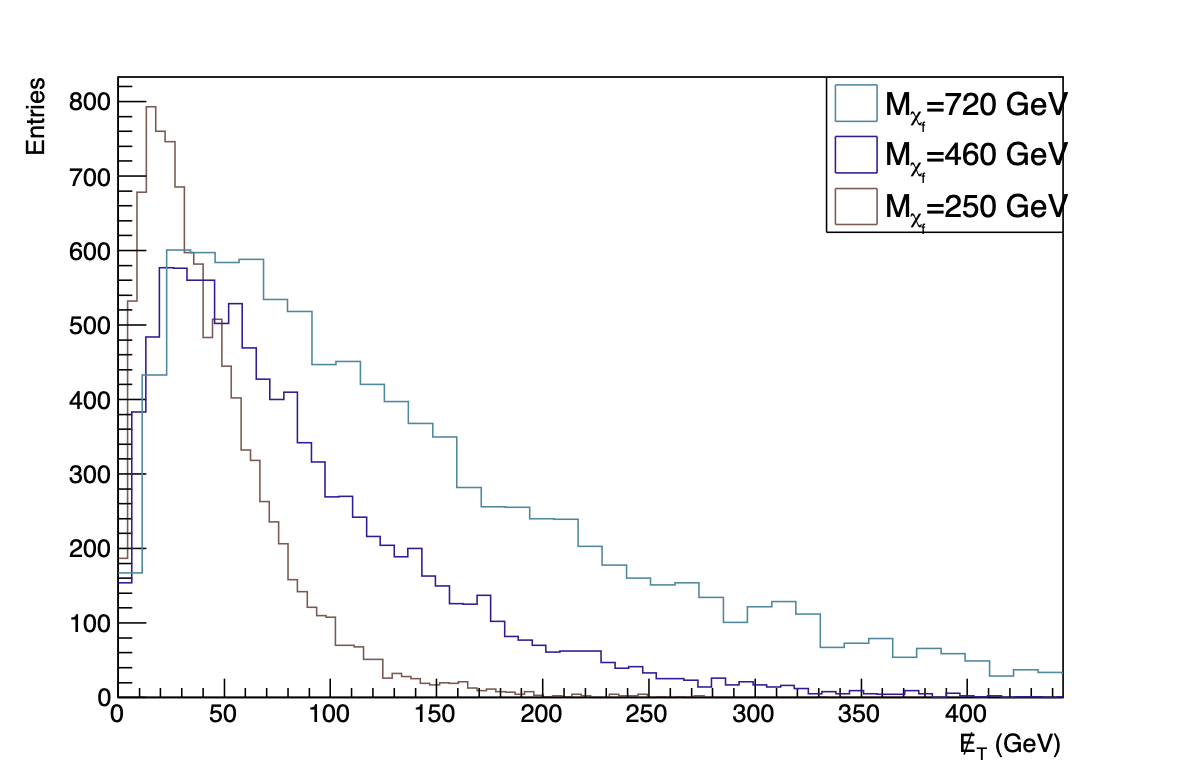}
\includegraphics[scale=0.40]{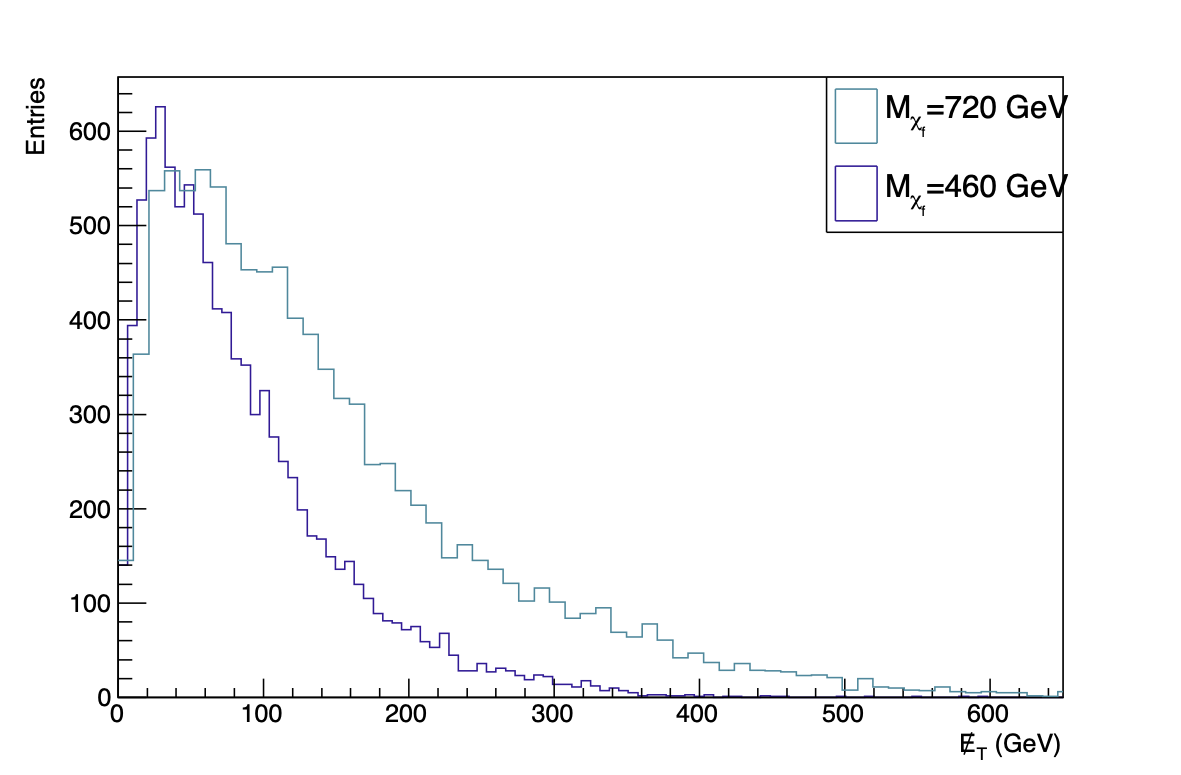}
\includegraphics[scale=0.40]{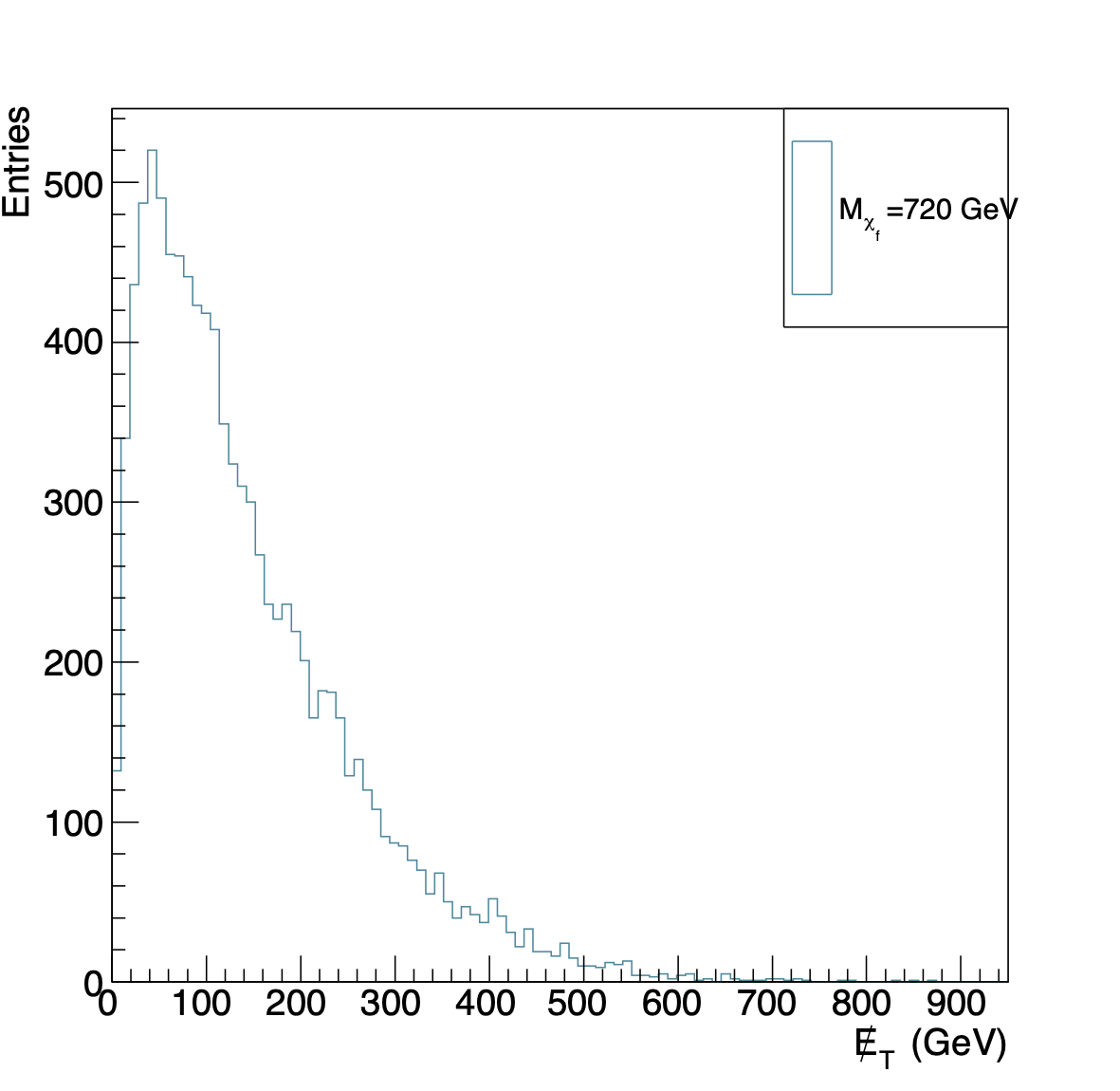}
	\caption{ $\slashed{E}_T$ for $p p \rightarrow \chi \bar{\chi} j j$ ,for 13 TeV LHC, for appropriate masses of dark matter candidate when $\Lambda=1500$ GeV, $\Lambda=2000$ GeV and $\Lambda=2500$ GeV .}
	\label{fig:dijet_met_lambda}
	\end{figure}
	
 In Fig. \ref{fig:dijet_met_lambda}, for dark matter pair production with two jets, the $\slashed{E}_T$ for $\Lambda=1500$ GeV, $\Lambda=2000$ GeV and $\Lambda=2500$ GeV are shown. As the mass increases, $\slashed{E}_T$ increases.
	
	\begin{figure}[H]
 \centering
\includegraphics[scale=0.60]{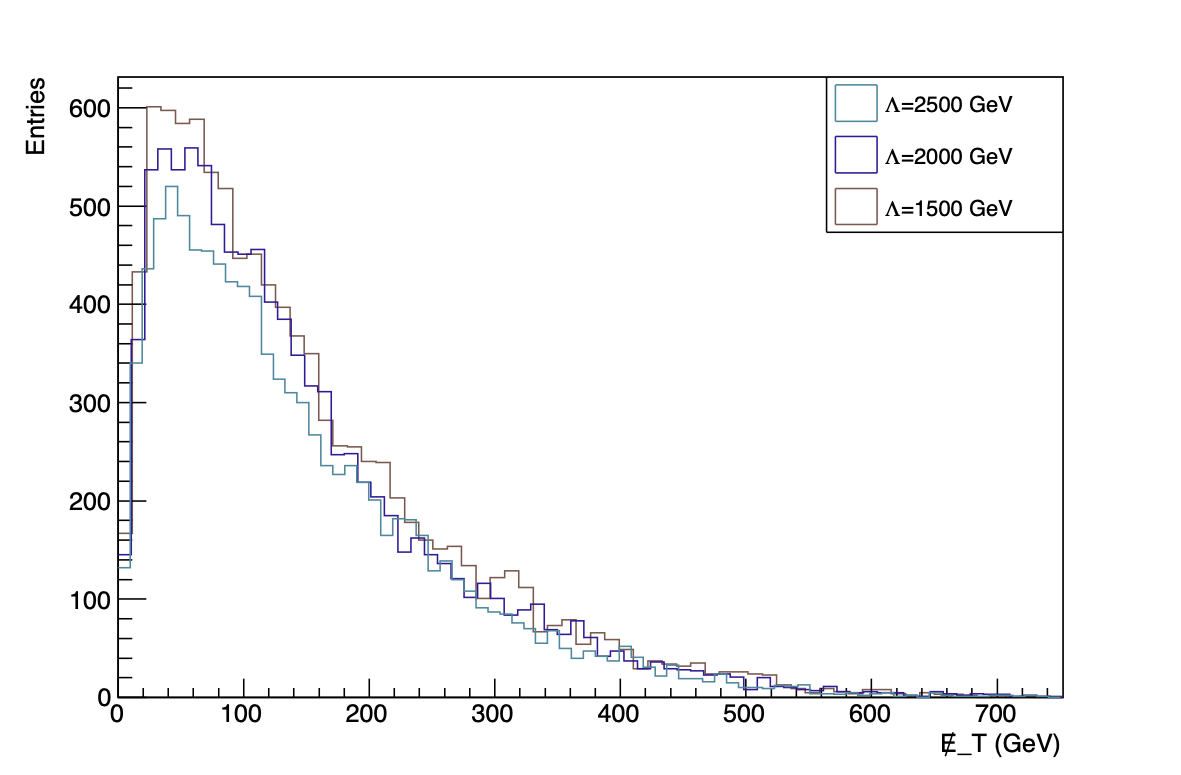} 
	\caption{$\slashed{E}_T$ for $M_{\chi}f=720$  GeV for $ pp\rightarrow \bar{\chi} \chi j j $ for $13$ TeV center of mass energy of LHC.}
	\label{fig:dijet_mxf720}
	\end{figure}
	
	Fig. \ref{fig:dijet_mxf720} shows characteristics of $\slashed{E}_T$ according to $\Lambda$, for the process of $pp\rightarrow \bar{\chi} \chi j j$. Because for the final state $\slashed{E}_T$ is originated from existence of dark matter, the behaviour of $\slashed{E}_T$,  does not change due to the $\Lambda$, and, it peaks on the same value. However, it is obvious that as $\Lambda$ increases since the effective vertices is supressed by cut-off value, the number of entries drop.

\section{Summary and Conclusion}
In this work, we studied six dimensional Effective Field Theory (EFT) of dark matter in details. These higher dimensional operators are introduced to  \textsc{FeynRules} in order to obtain UFO file.  Afterwords the generated UFO file is put into the \textsc{MadGraph5\_aMC@NLO} model file. By using \textsc{MadDM} tool of \textsc{MadGraph5\_aMC@NLO}, the parameter region of EFT was evaluated  according to the current proper relic density. In this DM model, we achieved the present dark matter candidate $\chi_f$ in the very simplistic way without the need of any new auxiliary particles or propagators. Thus, we proposed dark matter interactions with SM model particles directly via EFT vertices. After constraining model parameters by following the same step , we focused on dark matter pair production using \textsc{MadGraph5\_aMC@NLO} with standard cuts. The chained tools as we do construct here allow to study DM candidates interacting with SM via the higher dimensional operators from cosmos to collider. One can investigate more realistic parameter region of any kind of DM candidate in the frame of tool collection delicately integrated. It should be noted that the required condition $1/M^2 = g/\Lambda^2$ is now bilaterally available to ensure the validity of EFT theory. This helps to scan the validity region of the theory on the $(g,\Lambda^2)$-plane without any hesitation. In an effort to compare the current experimental results, for 13 TeV LHC, the analyses are made with 10000 events with standard cuts. For this comparision, as most common final states used in DM searches, monojet and dijet final states are investigated. When model parameters $\alpha_{\phi \chi (u\chi, d\chi, e \chi, q \chi, \ell \chi)}$ is set to 1, for $\Lambda=1500$ GeV, $M_{\chi_f}\geq 250$  GeV, for $\Lambda=2000$ GeV, $M_{\chi_f}\geq 460$ GeV, and for $\Lambda=2500$ GeV için $M_{\chi_f}\geq 720$ GeV values gives the current observed proper relic density region. By using this parameter set, $p_T$, $\eta$ and $\slashed{E}_T$ distributions are analysed. Results of this study are compatible with the latest results of ATLAS and CMS experiments and other previous studies. This confirms the reliability of our framework at least for fermionic DM candidate. Next steps will be to analyze the types of DM candidates given in the literature in the concept of EFT. Besides this study, we are planning to adapt the framework we developed, to perform EFT researches in the future-colliders.

\section*{Acknowledgments}

We would like express our gratest gratitude and thanks to \"Ozer  \"Ozdal and Prof. Dr. Taylan Yetkin for their fruitful discussion and helpful comments, and to  Prof. Dr. Benjamin Fuks for his help. Ayse El\c{c}{\.i}bo\u{g}a  Kuday is also supported by TUBITAK  2211/C Priority Areas of Domestic PhD Scholars\appendix

\section{Six-Dimensional Fermionic Effective Field Theory}
From the paper \cite{ref} the relation between coupling constants $g$'s and effective dark operators $\alpha$'s is:

 \begin{eqnarray}
	g^{u}_{L}=-\frac{1}{2}\alpha_{q\chi}, \qquad g^{u}_{R}=\frac{1}{2}\alpha_{u\chi} \nonumber
	\end{eqnarray}
	\begin{eqnarray}
	g^{d}_{L}=-\frac{1}{2}\alpha_{q \chi}, \qquad g^{d}_{R}=\frac{1}{2}\alpha_{d\chi} \nonumber
	\end{eqnarray}
	\begin{eqnarray}
	g^{e}_{L}=-\frac{1}{2}\alpha_{\ell \chi},\qquad g^{e}_{R}=\frac{1}{2}\alpha_{e\chi} \nonumber
	\end{eqnarray}
	\begin{eqnarray}
	g^{\nu}_{L}=-\frac{1}{2}\alpha_{\ell \chi}, \qquad g^{\nu}_{R}=0 \nonumber
	\end{eqnarray}

\section*{}

\end{document}